\documentclass[aps,prl,twocolumn,groupedaddress]{revtex4}
\usepackage{graphicx}
\usepackage{verbatim}
\usepackage{subfigure}
\usepackage[english]{babel}
\usepackage{siunitx}
\usepackage{pifont}
\usepackage{amsmath}
\usepackage{mathtools} 
\usepackage{upgreek}
\usepackage{multirow}

\begin{document}

\title{Benchmarking theory with an improved measurement of the ionization and dissociation energies of H$_2$}

\author{Nicolas~H{\"o}lsch$^1$, Maximilian~Beyer$^1\footnote{Present address: Department of Physics, Yale University, New Haven, CT 06511, USA}$, Edcel~J.~Salumbides$^2$, Kjeld~S.~E.~Eikema$^2$, Wim~Ubachs$^2$, Christian~Jungen$^3$ and Fr\'ed\'eric~Merkt$^{1,2}\footnote{Corresponding author; merkt@xuv.phys.chem.ethz.ch}$}

\affiliation{
	$^{1}$ Laboratorium f\"ur Physikalische Chemie,
	ETH-Z\"urich, 8093 Z\"urich, Switzerland\\
	$^{2}$ LaserLaB,
	Department of Physics and Astronomy, Vrije Universiteit, De
	Boelelaan 1081, 1081 HV Amsterdam, The Netherlands \\
	$^{3}$ Department of Physics and Astronomy, University College London, UK}

\date{\today}

\begin{abstract}
The dissociation energy of H$_2$ represents a benchmark quantity to test the accuracy of first-principles calculations. We present a new measurement of
the energy interval between the EF $^1\Sigma_g^+(v=0,N=1)$ state and the 54p1$_1$ Rydberg state of H$_2$. When combined with previously determined intervals, this new measurement leads to an improved value of the dissociation energy $D_0^{N=1}$ of ortho-H$_2$ that has, for the first time, reached a level of uncertainty that is three times smaller than the contribution of about 1 MHz resulting from the finite size of the proton. The new result of 35\,999.582\,834(11)~cm$^{-1}$ is in remarkable agreement with the theoretical result of 35\,999.582\,820(26)~cm$^{-1}$ obtained in calculations including high-order relativistic and quantum electrodynamics corrections, as reported in the companion article (M. Puchalski, J. Komasa, P. Czachorowski and K. Pachucki, submitted). This agreement resolves a recent discrepancy between experiment and theory that had hindered a possible use of the dissociation energy of H$_2$ in the context of the current controversy on the charge radius of the proton. 
\end{abstract}

\maketitle
The dissociation energy of molecular hydrogen, $D_0$(H$_2$), has been used as a benchmark quantity for first-principles quantum-mechanical calculations of molecular structure for more than a century. H$_2$ consists of two protons and two electrons and is the simplest molecule displaying all aspects of chemical binding. Whereas early calculations were concerned with explaining the nature of the chemical bond \cite{bohr13c,heitler27a,james33a}, the emphasis later shifted towards higher accuracy of the energy-level structure, requiring the consideration of nonadiabatic, relativistic and radiative contributions \cite{kolos60a,kolos63a,kolos68b,wolniewicz95c,bubin03a,piszczatowski09a,matyus12a,puchalski17a}.

These theoretical developments were accompanied and regularly challenged by experimental determinations of $D_0$(H$_2$) \cite{langmuir12a,witmer26a,beutler34a,herzberg61c,herzberg69a,stwalley70a,eyler93a,liu09b,cheng18a}. 
Periods of agreement between theory and experiment have alternated with periods of disagreement and debate. The reciprocal stimulation of theoretical and experimental work on the determination of $D_0$(H$_2$) has been a source of innovation. With its ups and downs and the related controversies, it has long reached epistemological significance \cite{primas84a,stoicheff01a}. 

In 2009, the experimental (36\,118.0696(4)~cm$^{-1}$) and theoretical (36\,118.0695(10)~cm$^{-1}$) values of $D_0^{N=0}$(H$_2$) reached unprecedented agreement at the level of the combined uncertainties of 30 MHz \cite{liu09b,piszczatowski09a}, apparently validating the treatment of the lowest-order ($\alpha^3$) QED correction and the one-loop term of the $\alpha^4$ correction, including several QED contributions that had not been considered for molecules until then. The insight that $D_0$(H$_2$) is a sensitive probe of the proton charge radius \cite{komasa11a,puchalski16a} stimulated further work. 

On the theoretical side, Pachucki, Komasa and coworkers have improved their calculations based on nonadiabatic perturbation theory \cite{pachucki14a,pachucki15a,pachucki16a,puchalski16a,puchalski17a}, significantly revised the 2009 result, and came to the unexpected conclusion that the excellent agreement of theoretical predictions with experimental $D_0$(H$_2$) values reached in 2009 was accidental, because of an underestimation of the contribution of nonadiabatic effects to the relativistic correction (see also Refs. \cite{wang18a,puchalski18a}). In the companion article, Puchalski {\it et al.} report on the theoretical progress, with a determination of the leading relativistic correction using the full nonadiabatic wave function \cite{puchalski19a}.

\begin{figure}[h]
	\includegraphics[trim=0cm 0cm 0cm 0cm, clip=true, width=1.0\columnwidth]{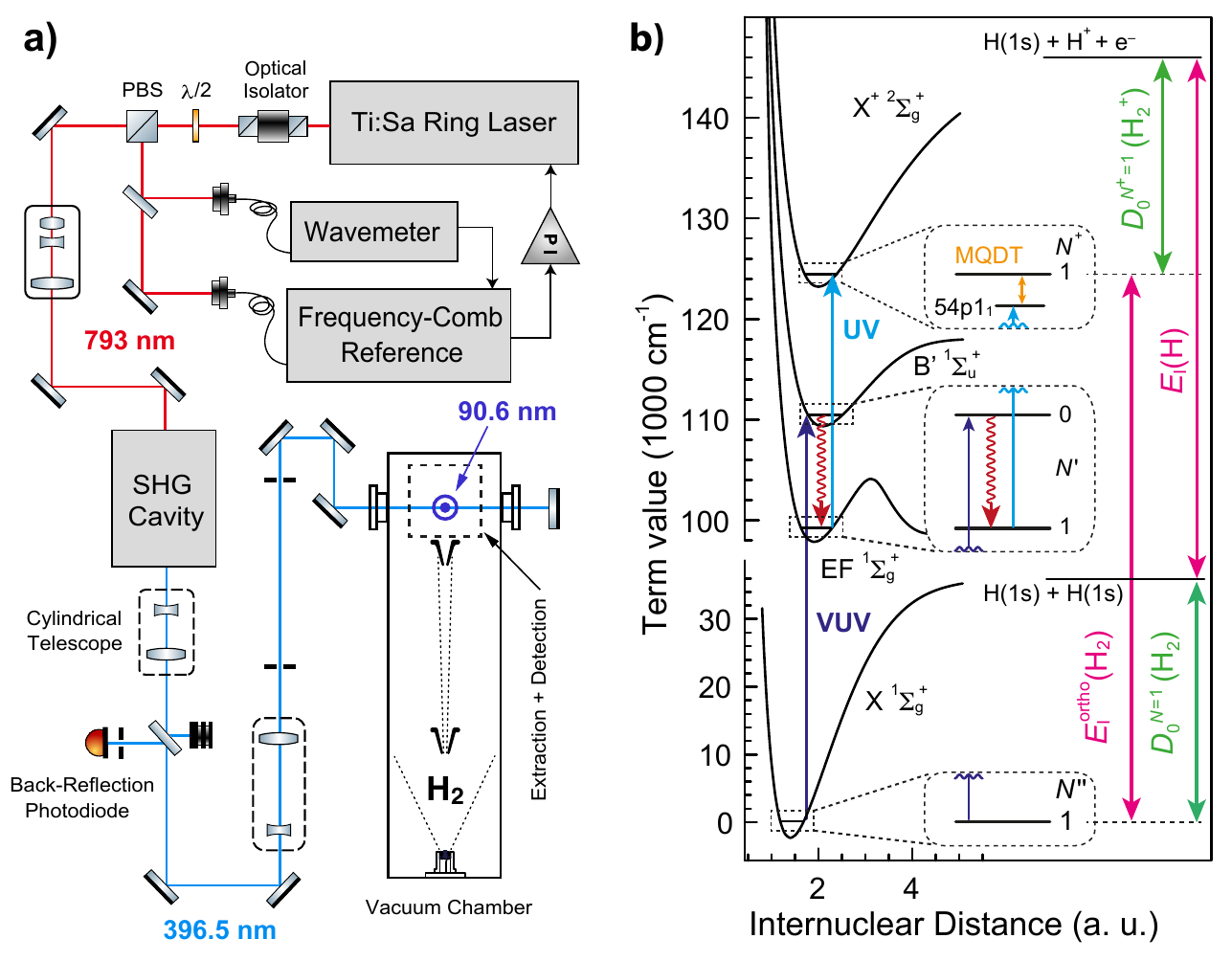}
	\caption{(a) Schematic diagram of the experimental setup. (b) Potential energy functions of the relevant electronic states of H$_2$ and H$_2^+$ and excitation scheme used to determine the ionization and dissociation energies of H$_2$ (see text for details).}
	\label{fig1}
\end{figure}

Recent experimental work has focused on the determination of the ionization energy $E_{\rm I}^{\rm ortho}({\rm H}_2)$ of ortho-H$_2$, from which the dissociation energy of ortho-H$_2$, $D_0^{N=1}({\rm H}_2)$, is obtained using (see Fig.~\ref{fig1}b)
\begin{equation}\label{cycle1}
D_0^{N=1}({\rm H}_2) = E_{\rm I}^{\rm ortho}({\rm H}_2) + D_0^{N^+=1}({\rm H}_2^+) - E_{\rm I}({\rm H})
\end{equation} 
and the very accurately known values of the ionization energy of the H atom ($E_{\rm I}({\rm H})$ \cite{mohr16a}) and of the dissociation energy of ortho-H$_2^+$, $D_0^{N^+=1}({\rm H}_2^+)$ \cite{korobov17a,korobova}. $E_{\rm I}^{\rm ortho}({\rm H}_2)$ is itself determined as the sum of energy intervals between the X $^1\Sigma_g^+(v=0,N=1)$ ground state and a selected low-$n$ Rydberg states, between this low-$n$ Rydberg state and a selected high-$n$ p Rydberg state, and the binding energy of the selected high-$n$ Rydberg state (see Ref.~\cite{sprecher11a} for details). 
In the 2009 determination, the selected low-$n$ and high-$n$ Rydberg states were the EF $^1\Sigma_g^+(v=0,N=1)$ and the 54p1$_1$ Rydberg state. To check the 2009 experimental result, $D_0$(H$_2$) was first determined by measuring the energy intervals between the X (0,1) and GK $^1\Sigma_g^+(v=1,N=1)$ states and between the GK(1,1) state and the 56p1$_1$ Rydberg state ~\cite{cheng18a}, after reevaluation of the binding energy of the $n$p1$_1$ Rydberg states \cite{sprecher14b}. 

In this Letter, we describe a new determination of $D_0$ through the EF(0,1) state with an absolute accuracy improved by a factor of 30 over the 2009 result. The new measurement is also 2.3 times more accurate than, and fully independent of, the measurement via the GK state mentioned above. 
The accuracy of the 2009 result was limited by the uncertainties arising from (1) the frequency chirps and spectral bandwidths of the pulsed lasers used to record spectra of the EF(0,1) - X(0,1) and 54p1$_1$ - EF(0,1) transitions, (2) ac-Stark shifts affecting the Doppler-free two-photon spectra of the EF(0,1) - X(0,1) transition, (3) dc-Stark shifts of the 54p1$_1$ - EF(0,1) transition resulting from ions generated in the measurement volume when preparing the EF(0,1) state by two-photon one-color excitation from the X(0,1) ground state, and (4) by the frequency calibration procedure, which relied on comparison with I$_2$ lines. 

These limitations have all been overcome: The effects of frequency chirps and ac-Stark shifts were eliminated by using a two-pulse Ramsey-comb method to determine the frequency of the EF(0,1) - X(0,1) transition \cite{altmann18a} and by using single-mode continuous-wave (cw) ultraviolet (UV) laser radiation to measure the 54p1$_1$ - EF(0,1) transition. When recording spectra of the 54p1$_1$ - EF(0,1) transition, the generation of ions was entirely suppressed by preparing the EF(0,1) state through single-photon excitation from the X(0,1) state to the B$^\prime(0,0)$ state, followed by spontaneous emission (SE):
\begin{equation}\label{eq:excitationGK}
\text{X}~^1\Sigma_g^+~(0,1)\xrightarrow{\rm VUV} \text{B}^\prime~^1\Sigma_u^+~(0,0)\xrightarrow{\rm SE} \text{EF}~^1\Sigma_g^+~(0,1).
\end{equation}
Finally, the relevant frequencies were all calibrated using frequency combs.  
The measurement of the X(0,1) - EF(0,1) interval by Ramsey-comb spectroscopy has been reported separately \cite{altmann18a}, and we describe here the measurement of the EF(0,1) - 54p1$_1$ interval, from which we derive $D_0$(H$_2$) with a 30-fold improved accuracy over the 2009 result \cite{liu09b}. 

The interval between the EF(0,1) state and the 54p$1_1(v^+=0,S=0,F=0-2)$ Rydberg state of ortho-H$_2$ was measured using the same molecular-beam apparatus and procedures as described in Ref.~\cite{beyer18a}, see Fig.~\ref{fig1}a. We refer to this work for details on, e.g., the compensation of the stray electric fields to better than 1~mV/cm and the shielding of external magnetic fields. The measurements were performed in a skimmed, pulsed supersonic beam of pure H$_2$ expanding into vacuum from a cryogenically cooled reservoir.

The pulsed vacuum-ultraviolet (VUV) radiation around 90.6~nm used to excite the ground-state molecules to the B$^\prime~^1\Sigma^+_u (v=0,N=0)$ state was produced in a four-wave mixing scheme as outlined in Ref.~\cite{beyer18a}. The lifetime of the B$^\prime~^1\Sigma^+_u$ state is of the order of 1~ns because of rapid spontaneous emission to the lower-lying EF~$^1\Sigma_g^+$ and X~$^1\Sigma_g^+$ states \cite{astashkevich15a}. The angular-momentum selection rule $\Delta N = \pm 1$ and Franck-Condon factors ensure that almost all molecules decaying to the EF state populate the $(v=0,N=1)$ rovibrational level. Further excitation from the EF(0,1) state to high-$n$ Rydberg states using the cw UV laser was detected by pulsed-field ionization (PFI), as described in Ref.~\cite{beyer18a}. The delay between the pulsed VUV radiation and PFI was set to 300~ns, i.e., longer than the lifetime of the EF state ($\tau({\rm EF})\approx200~\text{ns}$), to ensure that a maximum number of molecules could be excited to Rydberg states. 

\begin{figure}[h]
	\includegraphics[trim=0cm 0cm 0cm 0cm, clip=true, width=0.95\columnwidth]{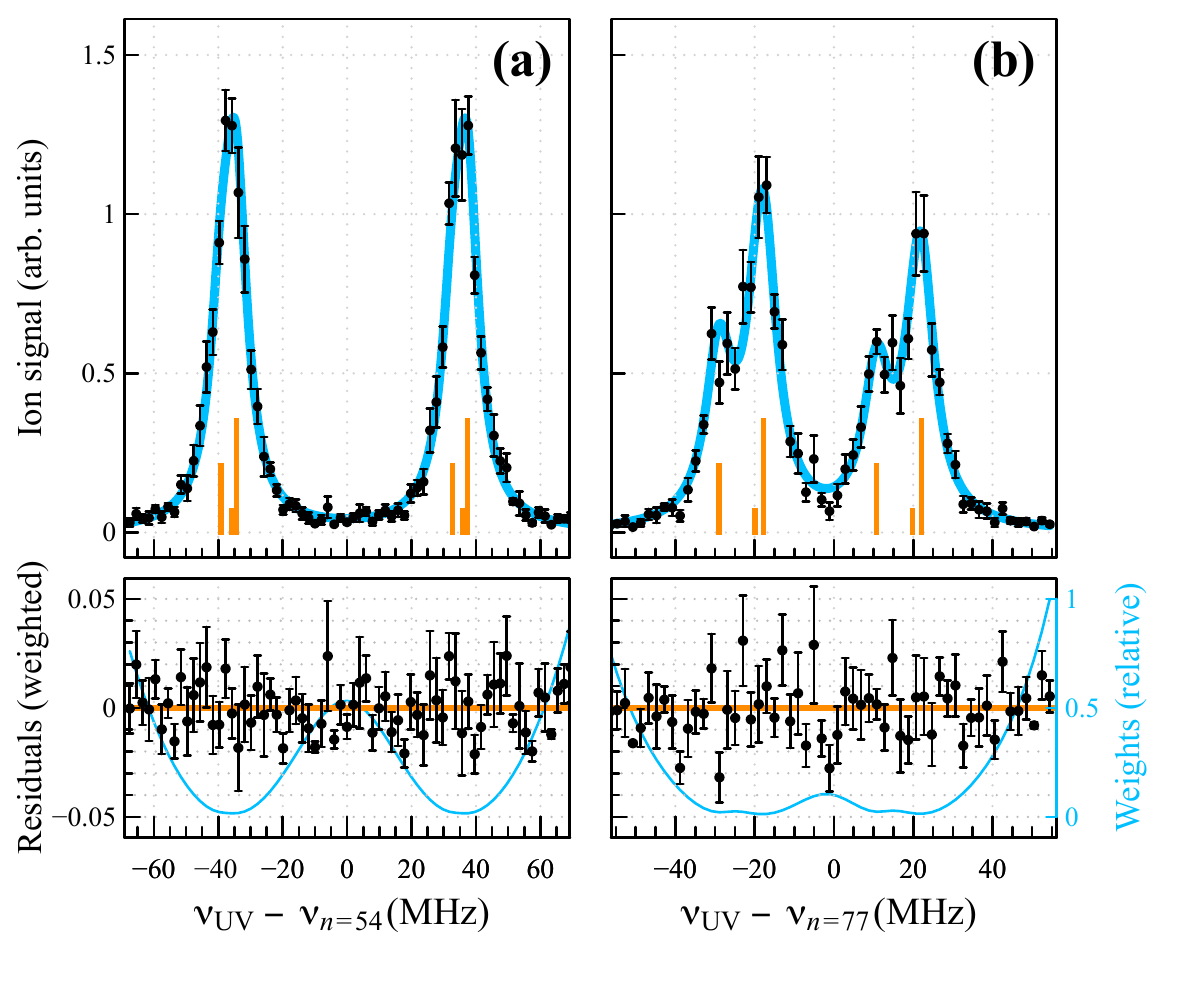}
	\caption{Upper panels: Spectra (black dots with error bars) of the 54p$1_1 \leftarrow {\rm EF}(0,1)$ (a) and 77p$1_1 \leftarrow {\rm EF}(0,1)$ (b) transitions and corresponding fits based on a Voigt line-shape model (blue traces), taking into account the hfs of the Rydberg states (orange stick spectra). Lower panels: Corresponding relative weights (blue) and weighted residuals.}
	\label{fig2}
\end{figure}

The cw UV radiation used to perform the excitation from the EF(0,1) level to the high-$n$ Rydberg states was generated by frequency-doubling the output of a single-mode (bandwidth $<50$~kHz) Ti:Sa ring laser in an external enhancement cavity containing an LBO crystal. The fundamental frequency of the Ti:Sa laser was calibrated with a frequency comb (accuracy better than 3~kHz) referenced to a 10-MHz GPS-disciplined Rb oscillator.  The UV laser beam crossed both the VUV laser beam and the molecular beam at  $\approx90^\circ$, was retro-reflected by a mirror, and crossed the H$_2$ sample again. Overlapping the forward and reflected UV-laser beams to better than 0.1~mrad and introducing a small deviation from 90$^\circ$ of the angle between UV and H$_2$ beam led to two well-separated Doppler components for each transition, from which the Doppler-free transition frequency could be determined as the average of the two peak centers (see Fig.~\ref{fig2} and corresponding discussion). 

\begin{figure}[h]
\includegraphics[trim=0cm 0cm 0cm 0cm, clip=true, width=0.99\columnwidth]{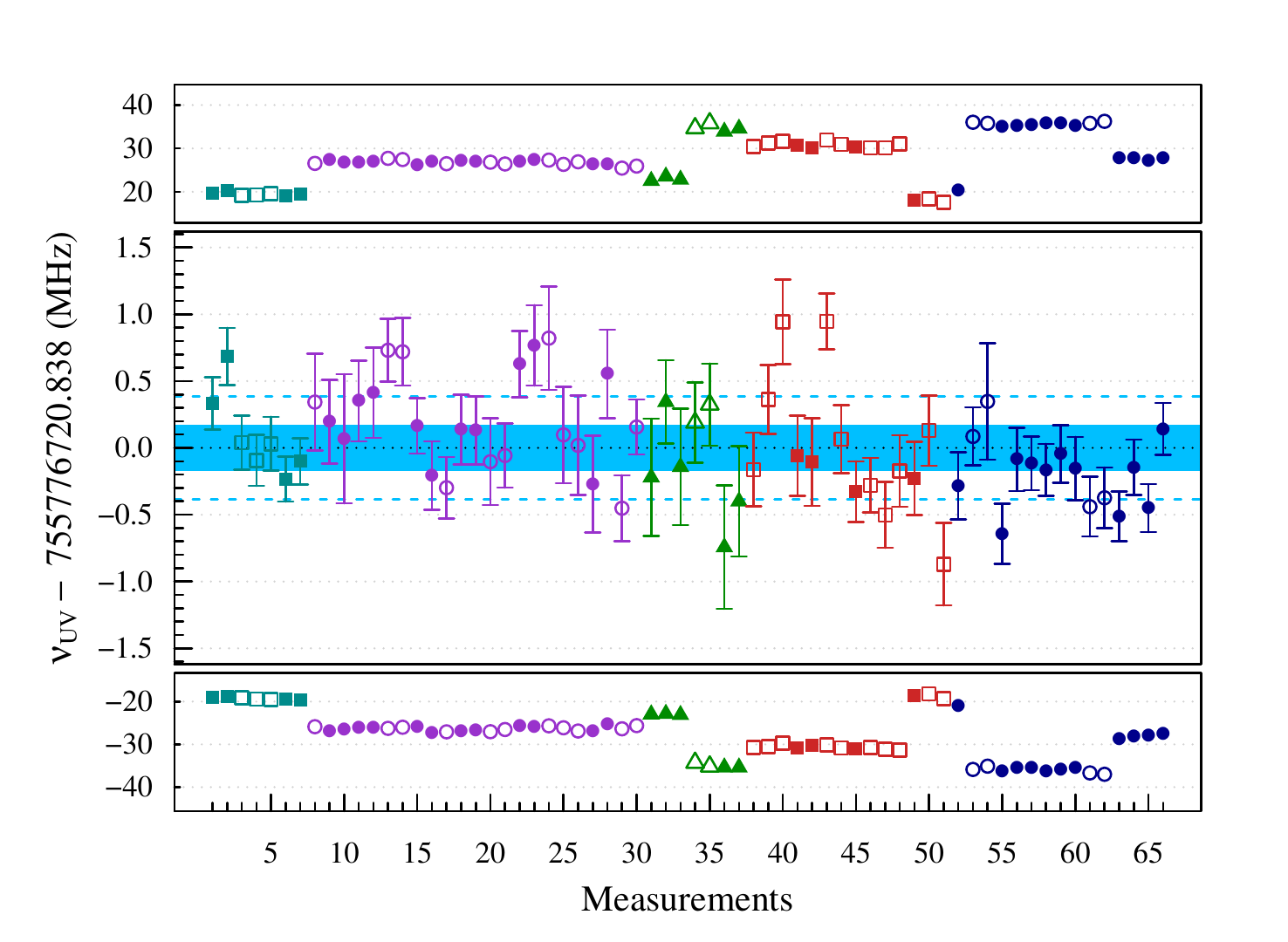}
\caption{Frequency of the 54p$1_1 \leftarrow {\rm EF}(0,1)$ transition of H$_2$ measured on five different days (indicated by different colors and symbols) and after regular realignment of the laser beam and its reflection (indicated by changes from full to open symbols). The valve temperature was 80~K. The top and bottom panels present the frequencies of the two Doppler components and the central panel displays their average value. The symbols and error bars represent the line positions obtained by fitting the center positions and their corresponding statistical uncertainties (one standard deviation), respectively. The dashed blue lines and the blue area give the standard deviations of the full data set and of the mean, respectively.}
\label{fig3}
\end{figure}

A telescope was used to place the focus of the UV beam onto the back-reflecting mirror and so ensure two identical Gaussian beams in the excitation region. The reflection angle was checked by monitoring the reflected beam through a 1-mm-diameter diaphragm located 8~m away from the reflection mirror. Complete realignment of the laser and molecular beams between measurements leads to a statistical uncertainty associated with the residual Doppler shift, instead of a systematic one. However, in the error budget, a systematic uncertainty of 200~kHz was included as upper limit for the effects of systematic misalignments. The cancellation of the first-order Doppler shift was verified independently using different angles (and therefore different Doppler shifts, see Fig.~\ref{fig3}) and by measuring the transition frequencies using fast and slow H$_2$ beams produced with the valve kept at room temperature and cooled to 80~K, respectively.

Fig.~\ref{fig2} displays typical spectra of the 54p$1_1 \leftarrow {\rm EF}(0,1)$ (a) and 77p$1_1 \leftarrow {\rm EF}(0,1)$ (b) transitions of H$_2$. Each transition in Fig.~\ref{fig2} splits into two Doppler components, corresponding to photoexcitation with the forward-propagating and reflected UV laser beams, as explained above.
The 54p$1_1 \leftarrow {\rm EF}(0,1)$ transition was selected because the binding energy and hyperfine structure (hfs) of the 54p$1_1$ Rydberg states are precisely known from previous studies combining millimeter-wave spectroscopy and multichannel quantum-defect theory (MQDT) \cite{osterwalder04a,sprecher14b}. The 77p$1_1 \leftarrow {\rm EF}(0,1)$ transition was used as reference and was measured after each realignment to detect possible drifts of the stray fields and of the UV-laser propagation axes with respect to the molecular-beam axis. Because the polarizability of Rydberg states scales as $n^7$, the 77p$1_1$ state is more than 10 times more sensitive to stray fields than the 54p$1_1$ state, making stray-field drifts of 1 mV/cm easily detectable. Such drifts would shift the frequency of the 54p$1_1 \leftarrow {\rm EF}(0,1)$ transition by less than 10 kHz. The hyperfine splittings of the $n$p$1_1$ series become larger with increasing $n$ value \cite{sprecher14b}. Consequently, the hfs of the 77p$1_1$ state could be partially resolved, which enabled us to verify experimentally that the intensities of the transitions to the three accessible ($F=0-2$) components are proportional to $2F+1$. Systematic uncertainties resulting from fits of the lineshapes with our lineshape model could thus be reduced to 100~kHz (see below and Refs.~\cite{beyer18a,hoelsch18a}). 

To determine the line positions, we fitted the lineshape model described in Ref.~\cite{beyer18a}, which consists, for each Doppler component, of a superposition of three line profiles corresponding to the three hyperfine components of the $n$p$1_1\,(F=0-2)$ Rydberg states, with intensities proportional to $2F+1$. For the 54p$1_1$ and 77p$1_1$ Rydberg states, we used the hyperfine splittings determined by millimeter-wave spectroscopy \cite{osterwalder04a} and MQDT calculations, respectively. Voigt profiles with a full width at half maximum of 9 MHz and a Lorentzian contribution of about 6 MHz were found to best reproduce the measured line profiles. The lineshape depends on the velocity distribution in the volume defined by the intersection of the VUV, UV and gas beams \cite{hoelsch18a,beyer18a}, with contributions from transit-time broadening and Doppler-broadening originating from the photon recoil of the B$^\prime\rightarrow$ EF spontaneous emission.
\begin{table}
	\caption{Error budget for the determination of the $54\text{p}1_1 \leftarrow \text{EF}(0,1)$ transition frequency}\label{table1}
	\begin{tabular}{lccc}
		\hline
		\textbf{Transition} 						& \multicolumn{2}{c}{$54\text{p}1_1 \leftarrow \text{EF}(0,1)$} \\
		Measured frequency 					& \multicolumn{2}{c}{755 776 720.84(18)~MHz}\\
		\hline
		 								& Correction & Uncertainty\\
		\hline
		DC Stark shift					 		&         & 10~kHz\\
		AC Stark shift					              	& & 5~kHz\\
		Zeeman shift 					              	& & 10~kHz\\
		Pressure shift					               	& & 1~kHz\\
		1st-order Doppler shift		       	& & 200 kHz \\
		2nd-order Doppler shift		             	& +8~kHz & 1~kHz\\
		Line-shape model						& & 100~kHz\\
		Hfs of EF(0,1)	  			& & 100~kHz \footnote{Estimated by multichannel quantum-defect theory in calculations of the type described in Ref.~\cite{osterwalder04a}} \\
		Photon-recoil shift		           	& -634~kHz &\\
		\hline	
		Systematic uncertainty		           	& & 250~kHz \\
		Final frequency 				           	& \multicolumn{2}{c}{755 776 720.21(18)$_{\text{stat}}(25)_{\text{sys}}\,{\text{MHz}}$} \\
		\hline
	\end{tabular}
\end{table}

Repeated measurements of these transitions revealed a high sensitivity of their frequencies to the alignment of the forward-propagating and reflected UV laser beams. Misalignments were detectable through an intensity imbalance between the two Doppler components. This effect turned out to be more pronounced than in our previous study of $n$p/f$\leftarrow {\rm GK}(1,1)$ transitions, an observation we attribute to the twice higher frequency of the $n$p$\leftarrow {\rm EF}(0,1)$ transitions and the resulting increased Doppler effect. In the final analysis of the data and after careful calibration of the effects of intentional, well-defined misalignments, we rejected all measurements associated with intensity ratios of the two Doppler components lying outside the range [0.8,1.25], and included a systematic uncertainty of 200 kHz (see above and Table~\ref{table1}).

Table~\ref{table1} also lists the other sources of systematic uncertainties considered in our analysis, which were estimated as explained in detail in Ref.~\cite{beyer18a}, and include uncertainties arising from DC and AC Stark shifts, Zeeman shifts, pressure shifts, Doppler shifts, and two contributions of each 100 kHz to account for uncertainties associated with the line-shape model and the unresolved (and unknown) hfs of the EF(0,1) state. The transitions frequencies were corrected by adding 8~kHz for the second-order Doppler shift and subtracting $634$~kHz for the photon-recoil shift, which is more than twice as large as the combined statistical and systematic uncertainty of 300 kHz.

The measurements used to determine the frequency of 54p$1_1\leftarrow {\rm EF}(0,1)$ transition were carried out at a valve temperature of 80~K and are depicted in Fig.~\ref{fig3}, which gives the central positions of the upper and lower Doppler components in the top and bottom panels, respectively,  and the average (Doppler-free and hyperfine-free) frequency with their statistical uncertainties (1$\sigma$) in the middle panel. The different colors and symbols indicate measurements carried out on different days and the sequence of full and open symbols indicate realignment of the laser beams. The dashed blue lines correspond to the standard deviation of the whole data set and the area shaded in blue the standard deviation of the mean. Adding the corrections listed in Table~\ref{table1} and combining all uncertainties in quadrature yields the value of {755\,776\,720.21(30)}~MHz ({25\,209.997\,785(10)}~cm$^{-1}$) for the 54p$1_1- {\rm EF}(0,1)$ interval.
\begin{table*}
	\caption{Overview of energy intervals used in the determination of the ionization and dissociation energies of H$_2$. See Fig. 1b for the relation between the different intervals.}\label{table2}
	\begin{tabular}{ccScc}
		\hline
		& Energy level interval & \multicolumn{1}{c}{Value (cm$^{-1}$)} & Uncertainty & Reference\\
		\hline\hline
		(1)& EF$(v=0,N=1)$ -- X$(v=0,N=1)$ 	&   99109.7312049(24)   & (73 kHz) & \cite{altmann18a}\footnote{Note that the first two columns of Table II of reference~\cite{altmann18a} unfortunately contain errors. The listed intervals in the first column add up to the ionization energy of ortho-H$_2$ instead of the dissociation energy of para-H$_2$, and the values given in the second column of the same table for the binding energy of the 54p1$_1$ Rydberg state and the dissociation energy of para-H$_2$ must be corrected to  37.509013(10) cm$^{-1}$ and 36 118.069 62(37) cm$^{-1}$, respectively.}	\\
		(2)& 54p$1_1(v^+=0,S=0,{\text{center}})$ -- EF$(v=0,N=1)$  &   25209.997785(10) 	 & (300 kHz)& \\
		(3)& X$^+(v^+=0,N^+=1,{\text{center}})$ -- 54p$1_1(v^+=0,S=0,{\text{center}})$ &   37.509013(5)	 &(150 kHz) & \cite{sprecher14b}	\\\hline
		(4)& $E^\text{ortho}_\text{I}$(H$_2$) = (1)+(2)+(3) 	 & 124357.238003(11)	 &(340 kHz)	& \\\hline
		
		(5)& $D^{N^+=0}_0$(H$_2^+$)		 & 21379.3502496(6) &(18 kHz)& \cite{korobov17a}	\\
		(6)& X$^+(v^+=0,N^+=1,{\text{center}})$ -- X$^+(v^+=0,N^+=0)$  & 58.2336750974(8) &(25 Hz) & \cite{korobova}	\\
		(7)& $D^{N^+=1}_0$(H$_2^+$)	= (5)-(6)	& 21321.1165745(6)	 &	(18 kHz) &		 \\
		
		(8)& $E_\text{I}$(H)   & 109678.77174307(10) & (3 kHz)& \cite{mohr16a} \\\hline
		
		(9)& $D^{N=1}_0$(H$_2$) = (4)+(7)-(8) 	&  35999.582834(11) & (340 kHz)& \\
		\hline\hline
	\end{tabular}
\end{table*}

Table~\ref{table2} provides the details of the determination of the ionization and dissociation energies of H$_2$ from the three intervals (entries (1), (2) and (3) in the table) linking the X(0,1) and X$^+$(0,1) ground states of ortho-H$_2$  and H$_2^+$ and corresponding to a value of $E_{\rm I}^{\rm ortho}({\rm H}_2)$ of {124\,357.238\,003(11)}~cm$^{-1}$. A value of {35\,999.582\,834(11)}~cm$^{-1}$ can be derived for $D_0^{N=1}({\rm H}_2)$ using Eq.~(\ref{cycle1}). The error budget in Table~\ref{table1} also applies to the 77p$1_1 - {\rm EF}(0,1)$ transition, with the exception of the uncertainty resulting from the dc Stark shift (100~kHz). A determination of $E_{\rm I}^{\rm ortho}({\rm H}_2)$ using the binding energy of the 77$p1_1$ state is in agreement with the results given in Table~\ref{table2}, proving the internal consistency of the MQDT analysis presented in Ref.~\cite{sprecher14b}.

Because of the very accurate value of the X-EF interval, our new result is more precise than the result of the measurement through the GK(0,1) state (35\,999.582\,894(25)~cm$^{-1}$ \cite{cheng18a}), from which it differs by about $2\sigma$. It is in agreement with the theoretical result (35\,999.582\,820(26)~cm$^{-1}$)  obtained by Puchalski {\it et al.} (see companion article \cite{puchalski19a}). This agreement between experiment and theory at the accuracy level of better than 1 MHz resolves the discrepancy noted in recent work \cite{puchalski17a} and may be regarded as unprecedented in molecular physics. The error margins within which theoretical and experimental values of $D_0^{N=1}$(H$_2$) agree are 30 times more stringent than in 2009. This agreement opens up the prospect of using $D_0$(H$_2$) to make a contribution to the solution of the proton-radius puzzle \cite{pohl10a}  as well as in the search or exclusion of fifth forces (see discussion in Ref.~\cite{salumbides13a}).
The experimental uncertainty of 340~kHz of the present result represents 30\% of the expected total contribution of 1 MHz to $D_0$ from the finite size of the proton \cite{puchalski16a}. The main sources of uncertainty of the present result come from the (unresolved) hfs of the EF(0,1) level, which affects both the X(0,1)-EF(0,1) and the EF(0,1) - 54p$1_1$ intervals and uncertainties associated with the residual first-order Doppler shift and the line shape model (see Table~\ref{table1}). These sources of uncertainties would be significantly reduced in a measurement in para-H$_2$, which should be the object of future efforts. In this context, theoretical work should consider the ionization energy of H$_2$ (see also \cite{puchalski19a}), which is the quantity we directly measure and which we obtain experimentally with a relative accuracy ($\Delta\nu/\nu$) of $9.1\cdot 10^{-11}$.

\begin{acknowledgments}
	FM, WU and KE acknowledge the European Research Council for ERC-Advanced grants under the European Union's Horizon 2020 research and innovation programme (grant agreements No 670168, No 743121 and No 695677). KE and WU acknowledge FOM/NWO for a program grant (16MYSTP). FM acknowledges the Swiss National Science Foundation (grant 200020-172620).
\end{acknowledgments}


\end{document}